\documentclass[a4paper,10pt]{llncs}
\usepackage[utf8]{inputenc}
\usepackage{url}
\usepackage{graphicx}

\title{Wages of wins:\thanks{The title ``Wages of wins'' was too good to pass up but is originally that of David Berri's book \cite{berri07wages}.} could an amateur\thanks{Someone without in-depth knowledge of the sport in question.} make money from match outcome predictions?}
\author{Albrecht Zimmermann}
\institute{Normandie Univ, UNICAEN, ENSICAEN, CNRS, GREYC, 14000 Caen, France}

\begin{document}

\maketitle

\begin{abstract}
Evaluating the accuracies of models for match outcome predictions is nice and well but in the end the real proof is in the money to be made by betting. To evaluate the question whether the models developed by us could be used easily to make money via sports betting, we evaluate three cases: NCAAB post-season, NBA season, and NFL season, and find that it is possible yet not without its pitfalls. In particular, we illustrate that high accuracy does not automatically equal high pay-out, by looking at the type of match-ups that are predicted correctly by different models.
\end{abstract}

\section{Introduction}

Using advanced sports analytics statistics and Machine Learning (or well-crafted mathematical) models, we can predict match outcomes for a variety of sports -- achieving predictive accuracies that are better than chance, a home-field advantage rule-of-thumb, or majority gut feeling. While this is an interesting (and in our opinion worthwhile) academic exercise, however, the question whether such work is actually useful becomes difficult to avoid. Or, to paraphrase a practitioner of sports betting: ``You should compare yourself to betting agencies and see whether you can make money!''.

In this work, we intend to do exactly this: using the example of three US sports attracting large betting volumes:
\begin{itemize}
 \item the main post-season tournament of university (NCAA) basketball -- \$ 9.2 billion bet (\$ 262 million legally),
 \item both regular and post-season of the National Basketball Association (NBA), and
 \item regular and post-season of the National Football League (NFL) -- Super Bowl alone \$ 4.1 billion bet (\$ 132 million legally)
\end{itemize}
we show not only predictive accuracies but also accumulated sports betting outcomes had we used their predictions to consistently place bets this year.


We find rather varying outcomes, and, in particular, that very similar accuracies can lead to strongly diverging monetary payoffs. To explore this phenomenon further, we
relate this to the way sports betting is handicapped.

In the following section, we discuss sports betting, and in particular how money-lines should be interpreted and are calibrated. We then discuss our experimental set-up before discussing hypothetical betting outcomes for the NCAAB post-season, the NBA season, and NFL season, respectively.

\section{Sports betting}

To understand the following discussions, it is necessary to understand money-lines offered by operators of sports betting services (\emph{sports books}), and to have some insight into how those money lines are derived.

US sports books offer two ways of betting on match outcomes: 
\begin{enumerate}
\item \emph{Over-under}, where bettors attempt to correctly foresee the difference between points score.
\item \emph{Money-line} betting, where bettors attempt to correctly divine the eventual winner of a match.
\end{enumerate}
Given that we have had weak results with trying to predict match scores in the past, we ignore the first setting for now, and focus on the second one, which allows us to relate binary predictions to monetary values. A money-line offered by a sports book for a particular match typically takes the form shown in the first row of Table \ref{tab:money-lines}.

\begin{table}
\begin{centering}
\begin{scriptsize}
\begin{tabular}{ccccc}
Match-up& Favorite (FAV)& Underdog (DOG)& FAV-Line& DOG-Line\\\hline
Detroit Pistons at Atlanta Hawks & ATL & DET &300 &240\\
Utah Jazz at Detroit Pistons&DET&UTH&110&-110\\
\end{tabular}
\end{scriptsize}
\caption{NBA money-line examples\label{tab:money-lines}}
\end{centering}
\end{table}

For each match, a probable winner (the Favorite) is identified, making the other team the probable loser (Underdog). The associated lines indicate the possible pay-out:
\begin{itemize}
 \item The FAV-Line indicates how much money one would \textbf{have to bet} to \emph{win} \$ 100.
 \item The DOG-Line indicates how much money one would \emph{win} if one \textbf{were to bet} \$ 100.
\end{itemize}
To make those two settings comparable, we can reformulate the FAV-Line since betting \$ 100 would net the bettor \$ $10000/$FAV-Line. For the first example given in Table \ref{tab:money-lines}, this means that Atlanta was considered the favorite and betting \$ 100 on them and winning would have paid out \$ 33.33. Detroit was expected to lose but if one had bet on them and they had defied predictions, one would have won \$ 240. 

Sports books do their best to calibrate those lines, trying to balance two attractions for bettors:
\begin{enumerate}
 \item Betting on the favorite is less risky and therefore has a higher chance to pay out.
 \item Betting on the underdog and winning will lead to a higher absolute pay-out.
\end{enumerate}
Ideally, a match's handicap attracts bettors in such way that the wins that the sports book needs to pay out are offset by the losses of those who bet on the other team (minus some profit for the sports book itself). This can be most clearly seen in the second example in Table \ref{tab:money-lines}, a so-called \emph{Pick 'em}. This is a match where the sports book operators do not have enough information to reliably predict one team as winning so betting on either one gives the same pay-off: \$ $10000/110 = 90.90$. Given a large enough number of bettors, one would expect that roughly half bets on either team and since the sports book pays out \$ 91 for every \$ 100 bet, it would stand to make a profit of 9\%.

\section{Experimental set-up}

Since we are going to use the same general set-up in the succeeding sections, we describe it here.

For each predictive setting, we have collected the money lines for all matches from \url{http://www.vegasinsider.com/}. The site lists the money-lines offered by the major sports books operating out of Las Vegas, Nevada, which occasionally differ slightly from each other. Additionally, money-lines vary with time, either due to the influx of new information (injuries, player arrests, coaches' announcements), or in reaction to bettors' behavior: too much interest in one team will lead to adjustment in favor of the other one. To avoid undue optimism when evaluating our predictors, we selected the most conservative line for each match. If a match is, for instance, listed once with FAV-Line=175, DOG-Line=155 and once with FAV-Line=165, DOG-Line=145, we will choose the latter since it would pay out less, no matter which prediction we make.

We use our models' predictions to select on which team to place the bet, and assume that we bet \$ 100 on every match in the time period. Correctly predicting a win by the favorite increases the model's winnings by \$ $10000/$FAV-Line, correctly predicting an underdog's win by \$ DOG-Line, and correctly predicting the winner of a Pick 'em by \$ 90.90. Incorrectly predicting a match outcome decreases winnings by \$ 100. For the sake of convenience, we predict matches, and tally up winnings, per day.

The preceding paragraph illustrates an important dynamic -- incorrectly predicting is always bad but not all correct predictions are equal:
\begin{itemize}
 \item Correctly predicting underdog wins is the most attractive option and depending on the money-line can balance out several incorrect predictions.
 \item Correctly predicting Pick 'ems still gives a relatively high pay-out.
 \item Correctly predicting favorite wins, on the other hand, needs to happen at a high rate to make up for incorrect predictions.
\end{itemize}

\section{NCAAB predictions (and bets)}

In our first setting, we consider the NCAAB post-season tournament, also referred to as ``March Madness'', for the interest and amount of sports betting it generates. This is the smallest of the settings we discuss since the tournament involved only 67 matches. We use the \emph{Adjusted Efficiencies} pioneered by Ken Pomeroy \cite{kenpom}, combined into a weighted average over the season, to encode teams, as well as season-level statistics such as the win percentage, margin of victory, point differential etc. For the full description of statistics, see \cite{zimmermann16sam}. We evaluate three classifiers: \emph{Na{\"i}ve Bayes} (NB), \emph{Multi-layer Perceptron} (ANN), and a simplified version of Ken Pomeroy's predictor based on the Pythagorean Expectation (KP). We referred to this classifier as ``simplified'' since we did not estimate the involved coefficients ourselves but based them on the discussions found on his blog. For the details of this classifier, see as well \cite{zimmermann16sam}. NB and ANN are used in their Weka \cite{weka} implementations, with default parameters, except that for NB \emph{Kernel estimator} is set to \emph{true}.

%

Before we discuss the performance of our classifiers, we need to establish the baseline. This means basing ourselves on the money-lines offered by sports books by assuming that we always follow the lead of the money-line. Concretely, if the team designated as favorite wins, we count this as a correct predictions for ``Vegas'', if the underdog wins, an incorrect one, with winnings accrued as described above. The main problem for this evaluation is posed by Pick 'ems: since the money lines give no indication but we would have to make a prediction, this amounts to flipping a coin for each Pick 'em. In the best case, we get each of those coin flips right, in the worst case, every single one wrong. Since the difference between getting a Pick 'em right and wrong amounts to \$ 190.90 per match (the lost gain + the \$ 100 bet), this leads to a large difference over the course of a season. Typically, we would assume to get half the coin flips right, which we report as expected accuracy and pay-out in Table \ref{tab:ncaab-vegas}.

\begin{table}
 \centering
 \begin{tabular}{c|c|c|c|c|c|c|c}
  \multicolumn{2}{c}{w/o Pick 'ems} & \multicolumn{6}{|c}{w/ Pick 'ems (5)}\\
  Accuracy & Pay-out & Best Acc. & Pay-out & Exp. Acc & Pay-out & Worst Acc. & Pay-out\\\hline
  0.7419 & 30.26 & 0.7611 & 484.76 & 0.7313& 7.51 & 0.6865& -469.73 
 \end{tabular}
\caption{Predictive accuracies and betting pay-outs for ``Vegas'' for the NCAAB post-season\label{tab:ncaab-vegas}}
\end{table}
We can see that always picking favorites would have gotten about 3/4 of the matches right, and paid out approximately \$ 30. Flipping coins on the Pick 'ems can lead to winnings of almost \$ 500 but also to losses of the same magnitude. Especially for so few (five) Pick 'ems, this is a very real risk.

\begin{table}
\centering
 \begin{tabular}{ccc}
  
 \begin{tabular}{l||c|c|c}
  Classifier & NB & ANN & KP\\\hline
  Accuracy & 0.6865 & 0.6417 & 0.7014\\
  Pay-out & 293.52 & -605.92 & -231.34\\
 \end{tabular}
 &\ \ \ \ &
 \begin{tabular}{l|c|c|c}
  Classifier & Favs & Dogs & Pick 'ems (of 5) \\\hline
  NB & 39 & 5 & 2 (0.4)\\
  ANN & 38 & 2 & 3 (0.6)\\
  KP & 43 & 0 & 4 (0.8)\\
 \end{tabular}
 \end{tabular}
 \caption{Predictive accuracies and betting pay-outs for three predictive models(left), Correct predictions by money-line characterization (right) for the NCAAB.\label{tab:ncaab-results}}
\end{table}

The results for the predictive models are shown on the left-hand side of Table \ref{tab:ncaab-results}. Two things are immediately noticeable: 1) the relative high predictive accuracies -- the KP model performs almost as well as the expected ``Vegas'' result, and 2) that this high accuracy does not translate into a high pay-out. Indeed, while Na{\"i}ve Bayes performs 1.5 percentage points worse than KP, it shows solid gains (better than using moneylines to bet only on favorites), while KP loses money.

We find some explanation for this phenomenon by looking at the right-hand side of Table \ref{tab:ncaab-results}. NB gets five upsets right, and even though KP is stronger in correctly predicting favorite wins and close Pick 'ems, this makes all the financial difference. The winnings curve for the different classifiers can be found in the appendix and shows that the winning behavior is rather erratic. Especially the ANN, which at some point posts winnings similar to the final outcome for NB, drops off into steep loss. But even the NB \emph{could} have returned twice of the final pay-out, a peak that is flanked by losses, however.

\section{NBA predictions (and bets)}

Our second setting concerns the NBA. We predicted matches for the 2016 regular and post-season, using NB, ANN, \emph{Random Forest} (RF),\footnote{Which we omitted for the NCAAB, as its accuracy is too weak.} as well as the simplified Ken Pomeroy model (KP). Teams were represented by the same statistics as for the NCAAB predictions. We did not predict the first two days of play since at that time a predictor would not have statistics for all teams. 
As in the preceding section, we need to establish the baseline, shown in Table \ref{tab:nba-vegas}.

\begin{table}[ht]
\centering
 \begin{tabular}{c|c|c|c|c|c|c|c}
  \multicolumn{8}{c}{Regular + post-season}\\
 \multicolumn{2}{c}{w/o Pick 'ems} & \multicolumn{6}{|c}{w/ Pick 'ems (115)}\\
  Accuracy & Pay-out & Best Acc. & Pay-out & Exp. Acc & Pay-out & Worst Acc. & Pay-out\\\hline
  0.7121 & -2374.16 & 0.7375 & 9125.84 & 0.6937 & -1857.3 & 0.6492 & -12828.81\\
 \end{tabular}
 \caption{Predictive accuracies and betting pay-outs for ``Vegas'' for the NBA\label{tab:nba-vegas}}
\end{table}

Following the money line over the course of the entire season, while ignoring the Pick 'ems, would lead to a very respectable accuracy but also to a monetary loss. At {\raise.17ex\hbox{$\scriptstyle\sim$}}1200 matches, the loss per match is only about \$ 1 yet over the course of the season this accrues. Getting half the Pick 'ems right does of course not improve this, even though the accuracy would stay high.

\begin{figure}[ht]
\centering
 \includegraphics[width=\linewidth]{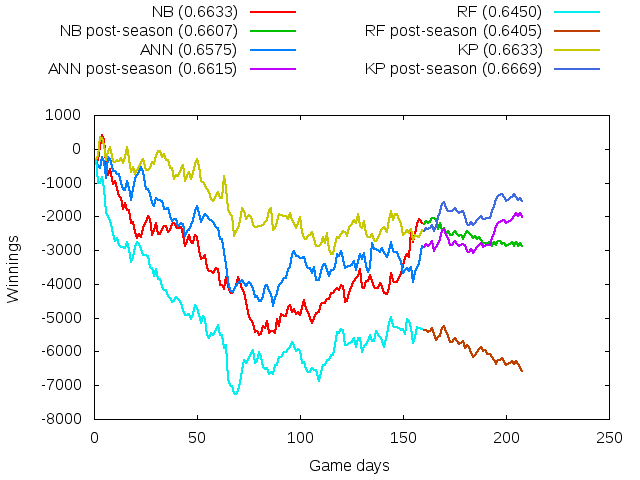}
 \caption{Classifier winnings over the course of the season, NBA\label{nba-winnings}}
\end{figure}

Figure \ref{nba-winnings} plots the development of the different classifiers' winnings over the course of the season, the legend is annotated with predictive accuracies. None of them show a net positive payout, and with the exception of the KP model, they all drop rather low. Notably, they all recover to a certain degree, however, meaning that one could win money if one could determine \emph{when to start betting}. Plots showing the difference between the trough and the best result, and its magnitude, can be found in the appendix (Figures \ref{nba-winnings-nb}--\ref{nba-winnings-kp}). For the ANN and KP, the best result is in the post-season, for NB and the RF in the regular season, even though NB and KP have the same regular season accuracy.
Table \ref{tab:nba-picks} shows why: while KP strongly outperforms NB in getting favorites right (as for the NCAAB), it underperforms when it comes to Pick 'ems. Pick 'ems are clearly the most difficult matches to predict, and with KP combining three estimated influences -- adjusted efficiencies, the coefficient in the Pythagorean Expectation, and the home-court adjustment -- small errors can spiral. The trough-peak difference aligns with the amount of underdog/pick 'em predictions.

\begin{table}
\begin{centering}
 \begin{tabular}{l|c|c|c||c|c|c}
 & \multicolumn{3}{c|}{Regular season}&\multicolumn{3}{|c}{Post-season}\\
  Classifier & Favs & Dogs & Pick 'ems (109)&Favs & Dogs & Pick 'ems (6)\\\hline
  NB & 691 & 57 & 48 (0.44)&49 &5&1 (0.16)\\
  ANN & 707 & 60 & 22 (0.20)&57&6&0\\
  RF & 685 & 61 & 28 (0.26) &47&4&0\\
  KP & 725 & 59 & 12 (0.11) &58&5&0\\
 \end{tabular}
 \caption{Correct predictions by money-line characterization for the NBA\label{tab:nba-picks}}
 \end{centering}
\end{table}


\section{NFL predictions (and bets)}

This season marked our first attempt at NFL predictions. As for basketball, the main question to answer concerns team representations. In basketball matches, individual events are \emph{possessions} that lead to either points, or a number of possibly possession-changing events. In American Football, on the other hand, individual events are \emph{Downs} and their outcome is mainly measured in \emph{yards gained} (or lost). While the more or less discrete results in basketball can be read off the final box score, the fluctuation of yards in football is less well captured. 

To address this, Football Outsiders have proposed \emph{Defense-adjusted Value Over Average} and \emph{Defense-adjusted Yards Above Replacement} \cite{foDVOA}, both of which consider the outcome of each down in relation to the league-wide average against a particular \emph{defensive alignment}. Since this requires access to and work with play-by-play statistics, we forwent this approach and instead evaluated several other statistics over past seasons:
\begin{itemize}
\item Basic Averages -- all the statistics available from a typical box score at \cite{footballReference} under "team stats", normalized for 65 possessions, and averaged in a weighted manner (recent games have more weight), both offensively (scored/gained/ committed) and defensively (allowed/caused). This follows similar reasoning as possession-based normalizing and averaging in basketball.
\item Opp. Averages -- same as above but for the opponents that have been played so far. This is supposed to help gauge the competition.
\item Adjusted Averages -- certain offensive and defensive statistics  adjusted by mirror statistics of the respective opponents. That is basically the same idea as Ken Pomeroys adjusted efficiencies \cite{kenpom}.
\item SRS -- the "simple rating system" information (SRS, SoS) as described at \cite{SRS}, with the difference that the averaging is weighted, so not divided by number of matches.
\end{itemize}
Page limitations prevent us from showing the full results of the evaluation here. We intend to write this down formally in the future but for the time being, the details can be found at \cite{sdmaz15NFL-representations}. After additional evaluation during the season, we settled on using Basic+Opponents' Averages for the NB, and Adjusted Averages for ANN and RF. We also evaluated the SRS. We did not predict the first week's matches since for those matches, we do not have statistics for the teams at that time.
Again, we need to establish the baseline (see Table \ref{tab:nfl-vegas}).

\begin{table}[ht]
\centering
 \begin{tabular}{c|c|c|c|c|c|c|c}
  \multicolumn{8}{c}{Regular + post-season}\\
 \multicolumn{2}{c}{w/o Pick 'ems} & \multicolumn{6}{|c}{w/ Pick 'ems (29)}\\
  Accuracy & Pay-out & Best Acc. & Pay-out & Exp. Acc & Pay-out & Worst Acc. & Pay-out\\\hline
  0.6441 & -1215.69 & 0.6852 & 1420.68 & 0.6294 & -1251.92 & 0.5697 & -4115.42
 \end{tabular}
 \caption{Predictive accuracies and betting pay-outs for ``Vegas'' for the NFL\label{tab:nfl-vegas}}
\end{table}

The baseline again shows consistent behavior: the accuracy is relatively high but if one follows money-line predictions one loses -- not much per individual game but quite a bit in the aggregate.

\begin{figure}[h]
\centering
 \includegraphics[width=\linewidth]{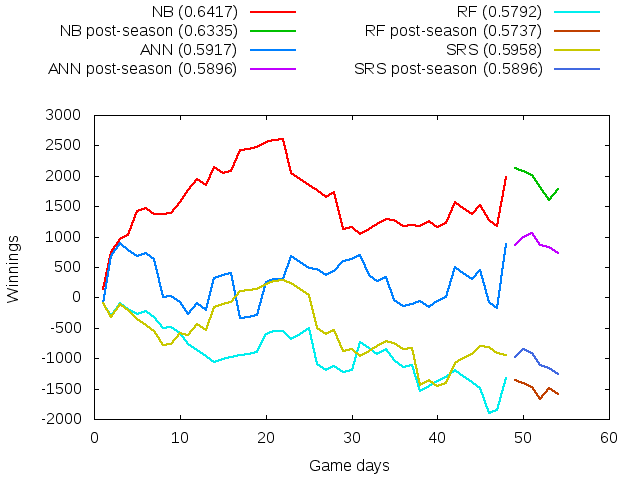}
 \caption{Classifier winnings over the course of the season, NFL\label{nfl-winnings}}
\end{figure}


The results of the predictors, shown in Figure \ref{nfl-winnings}, are very interesting.
The first thing to notice is that NB, using rather straight-forward statistics, achieves comparative accuracy to ``Vegas'' and a much better pay-out. In fact, its pay-out is better than that for the best-case ``Vegas''-scenario in the regular season. Even the ANN, with much lower accuracy, achieves a good pay-out. Additionally, we again see the influence of which picks to predict correctly at play: even though ANN and SRS have very similar accuracies, betting according to SRS would be a clear loss, and the difference can be explained by the fact that the ANN trades off accuracy on favorites against accuracy on underdogs (Table \ref{tab:nfl-picks}).

\begin{table}
\begin{centering}
 \begin{tabular}{l|c|c|c||c|c|c}
 & \multicolumn{3}{c|}{Regular season}&\multicolumn{3}{|c}{Post-season}\\
  Classifier & Favs & Dogs & Pick 'ems (28)&Favs & Dogs & Pick 'ems (1)\\\hline
  NB & 115 & 25 & 14 (0.5)&4 &1&0\\
  ANN & 98 & 29 & 15 (0.54) &5&0&1 (1.0)\\
  RF & 107 & 16 & 16 (0.57) & 4&1&0\\
  SRS & 111 & 18 & 14 (0.5) &4&0&1 (1.0)\\
 \end{tabular}
 \caption{Correct predictions by money-line characterization for the NFL\label{tab:nfl-picks}}
 \end{centering}
\end{table}

A final monetary realization is that each predictor reaches a high point that comes before the end of the season. In fact, following NB all to the end of the regular season would mean forfeiting more than \$ 600, with losses for all models in the post-season. While for NBA predictions it seems to be important to know when to get \emph{in}, for the NFL is important to know when to get \emph{out} -- a decision that might be slightly easier to make.

\section{Conclusion and outlook}

The answer to the question posed in the title of the paper is a definitive ``Maybe!''. Once a model has been established, it can be used to place bets in a straight-forward manner. 
However, the NCAAB post-season contains few matches, leading to rather volatile pay-out.
In the NBA, one can win but only after figuring out when to start betting.
In the NFL results, finally, straight-forward use could indeed lead to a decent pay-off (admittedly, not attractive to professional gamblers), especially if one stops early enough. In all cases, the safest model seems to be a Na{\"i}ve Bayes predictor.

We have tried to show one of the aspects that make a predictive model more or less well-suited for sports betting, by considering what kind of matches models predict well. In particular, a model that is not very strong in correctly predicting favorites but gets a large amount of Pick 'ems correct, or even better matches won by underdogs, would be a particular attractive tool, even if its straight-up accuracy is not impressive.

We intend to explore this question further by relating models' performance to evaluations based on lift-charts and ROC-like discussions. We have the data needed for this exploration already available (and plotted) but page constraints prevent us from discussing it in this work. The final goal would of course be to shift the training of predictive models: away from maximizing predictive accuracy and towards maximizing pay-outs, which means getting border-line cases right instead of easy ones. A different direction consists of proposing which matches (not) to bet on and/or how much to bet, as has been done in \cite{DBLP:conf/scai/Langseth13,snyder2013actually} for soccer. Possible approaches include leveraging game theoretic approaches or reinforcement learning.\footnote{We thank the reviewers for this suggestion.}

\bibliography{../bibliographie}
\bibliographystyle{splncs03}

\appendix

\section{Vegas results}

\begin{table}[ht]
\centering
 \begin{tabular}{c|c|c|c|c|c|c|c}
 \multicolumn{8}{c}{Regular season}\\
 \multicolumn{2}{c}{w/o Pick 'ems} & \multicolumn{6}{|c}{w/ Pick 'ems}\\
  Accuracy & Pay-out & Best Acc. & Pay-out & Exp. Acc & Pay-out & Worst Acc. & Pay-out\\\hline
  0.7096 & -1502,00 & 0.7356 & 8407.09 & 0,6904 & -1983,14 & 0.6461 & -12402\\
  \multicolumn{8}{c}{Regular + post-season}\\
 \multicolumn{2}{c}{w/o Pick 'ems} & \multicolumn{6}{|c}{w/ Pick 'ems}\\
  Accuracy & Pay-out & Best Acc. & Pay-out & Exp. Acc & Pay-out & Worst Acc. & Pay-out\\\hline
  0,7121 & -2374,16 & 0,7375 & 9125.84 & 0,6937 & -1857,3 & 0,6492 & -12828,81\\
 \end{tabular}
 \caption{Predictive accuracies and betting pay-outs for ``Vegas'' for the NBA\label{tab:nba-vegas-detailed}}
\end{table}

\begin{table}[ht]
\centering
 \begin{tabular}{c|c|c|c|c|c|c|c}
 \multicolumn{8}{c}{Regular season}\\
 \multicolumn{2}{c}{w/o Pick 'ems} & \multicolumn{6}{|c}{w/ Pick 'ems}\\
  Accuracy & Pay-out & Best Acc. & Pay-out & Exp. Acc & Pay-out & Worst Acc. & Pay-out\\\hline
  0,6367 & -1443,45 & 0,6791 & 1102.49 & 0,6208 & -1570,11 & 0,6375 & -4242,71\\
  \multicolumn{8}{c}{Regular + post-season}\\
 \multicolumn{2}{c}{w/o Pick 'ems} & \multicolumn{6}{|c}{w/ Pick 'ems}\\
  Accuracy & Pay-out & Best Acc. & Pay-out & Exp. Acc & Pay-out & Worst Acc. & Pay-out\\\hline
  0,6441 & -1215,69 & 0,6852 & 1420.68 & 0,6294 & -1251,92 & 0,5697 & -4115,42
 \end{tabular}
 \caption{Predictive accuracies and betting pay-outs for ``Vegas'' for the NFL\label{tab:nfl-vegas-detailed}}
\end{table}
\newpage
\section{NCAAB winning curves}

\begin{figure}[h]
\centering
 \includegraphics[width=0.8\linewidth]{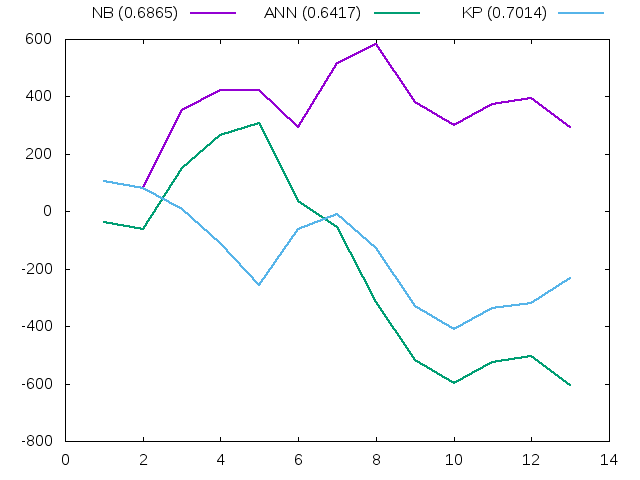}
 \caption{Classifier winnings over the course of the season, NCAAB\label{ncaab-winnings}}
\end{figure}
\newpage
\section{Detailed NBA winning curves}

\begin{figure}[th]
\centering
 \includegraphics[width=0.6\linewidth]{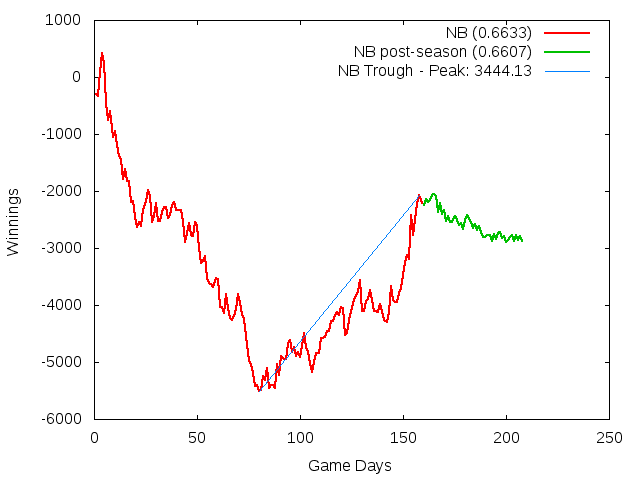}
 \caption{Trough to peak winnings for the NBA season, NB\label{nba-winnings-nb}}
\end{figure}

\begin{figure}[h]
\centering
 \includegraphics[width=0.6\linewidth]{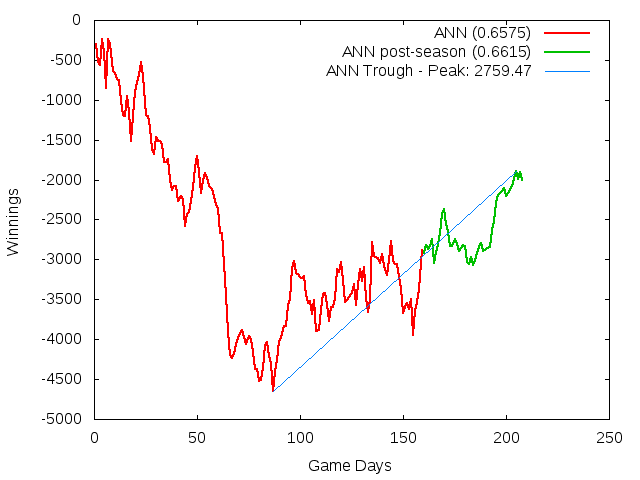}
 \caption{Trough to peak winnings for the NBA season, ANN\label{nba-winnings-ann}}
\end{figure}

\begin{figure}[th]
\centering
 \includegraphics[width=0.6\linewidth]{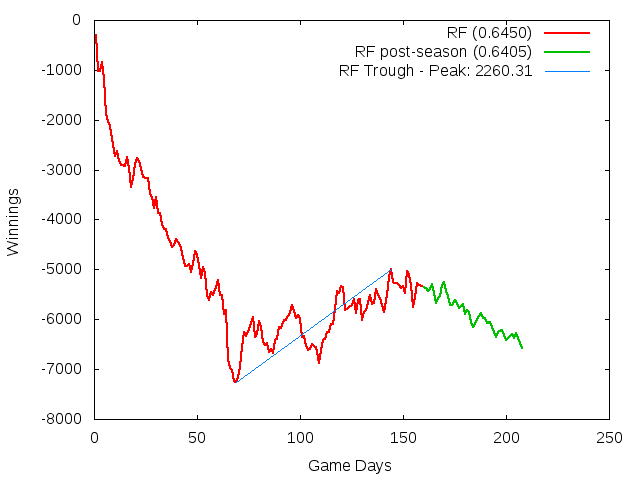}
 \caption{Trough to peak winnings for the NBA season, RF\label{nba-winnings-rf}}
\end{figure}

\begin{figure}[h]
\centering
 \includegraphics[width=0.6\linewidth]{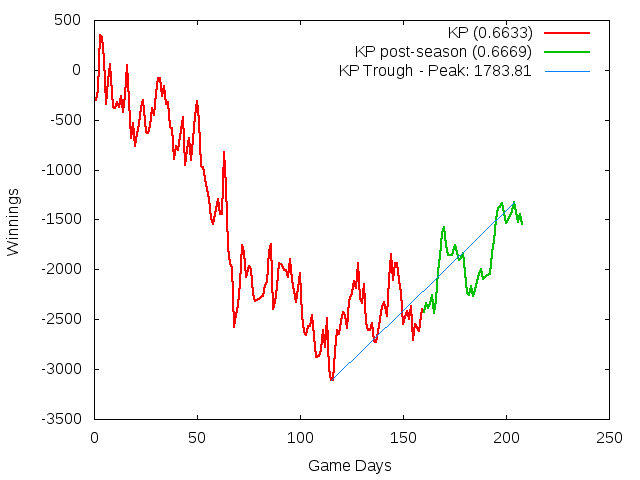}
 \caption{Trough to peak winnings for the NBA season, KP\label{nba-winnings-kp}}
\end{figure}

\end{document}